
\input harvmac
\def\sixth{\textstyle{1\over 6}}\def\fourth{\textstyle{1\over 4}}
\def\mnl{{\mu\nu\lambda}}
\def\weakly{\approx}
\def\PB#1#2{\big\{#1, #2\big\}}
\def\pb{Poisson bracket}
\def\eijk{\epsilon_{ijk}}\def\eklm{\epsilon_{klm}}
\def\wave{\mbox{0.1}{0.1}}
\def\DB#1#2{\big\{#1, #2\big\}_D}
\def\bx{{\bf x}}\def\by{{\bf y}}
\noblackbox
\lref\ablii{T. J. Allen, M. J. Bowick and A. Lahiri,
{\sl Mod. Phys. Lett.} {\bf A6} (1991) 559.}
\lref\mw{
J. Minahan and R. Warner, {\sl
Stuckelberg Revisited}, University of Florida preprint UFIFT-HEP-89-15
(unpublished).}
\lref\tja{T. J. Allen, ``Duality and the Vacuum'', {\sl Nucl. Phys. B (to be
published)}.}
\lref\bal{A. P. Balachandran and P. Teotonio-Sobrinho, ``The
   Edge states of the BF system and the London
   equations'', {\sl Int. J. Mod. Phys. A (to be published)}.}
\lref\kyl{K. Lee, ``The Dual Formulation of Cosmic Strings and Vortices'',
Columbia U. preprint CU-TP-588, hep-th: 9301102.}
\lref\davshe{R. L. Davis and E. P. S. Shellard, {\sl Phys. Lett.} {\bf
B214} (1988) 219; {\sl Phys. Rev. Lett.} {\bf 63} (1989) 2021.}
\lref\davis{R. L. Davis, {\sl Phys. Rev.} {\bf D40} (1989) 4033; {\sl Mod.
Phys. Lett.} {\bf A5} (1990) 955.}
\lref\op{V. I. Ogievetskii and I. V. Polubarinov, {\sl Sov. J. Nucl. Phys.}
({\sl Iad. Fiz.}) {\bf 4} (1967) 156.}
\lref\des{S. Deser, {\sl Phys. Rev.} {\bf 187} (1969) 1931.}
\lref\kr{M. Kalb and P.
Ramond,
{\sl Phys. Rev.} {\bf D9} (1974) 2273.}
\lref\rwit{R. Rohm and E. Witten, {\sl Ann.
Phys.} {\bf 170} (1986) 454.}
\lref\vilen{A. Vilenkin and T. Vachaspati, {\sl Phys. Rev.} {\bf D35}
(1987) 1138.}
\lref\bghhs{M. J.  Bowick {\sl et al.}, {\sl Phys. Rev. Lett.} {\bf
61} (1988) 2823.}
\lref\abli{T. J. Allen, M. J. Bowick and A.
Lahiri, {\sl Phys. Lett.} {\bf B237} (1989) 47.}
\lref\cdno{Bruce A. Campbell,{\sl et al.}, {\sl Phys. Lett.} {\bf B251}
(1990) 34; {\sl Nucl. Phys.} {\bf B351} (1991) 778.}
\lref\burg{C. P. Burgess and A. Kshirsagar, {\sl Nucl.Phys.} {\bf B324} (1989)
157\semi J. M. Molera and B. A. Ovrut, {\sl Phys. Rev.} {\bf D40} (1989) 1146.}
\lref\dirac{P. A. M. Dirac, {\sl Can. J. Math.} {\bf 2} (1950) 129; {\sl Proc.
Roy. Soc.} {\bf A246} (1968) 326\semi
A. J. Hanson, T. Regge and C. Teitelboim, {\sl Constrained Hamiltonian
Systems}, Accad. Naz. dei Lincei, Rome, 1976.}
\lref\kaul{R. K. Kaul, {\sl Phys. Rev.} {\bf D18} (1978) 1127.}
\lref\cpw{S. Coleman, J. Preskill and F. Wilczek, {\sl Nucl. Phys.} {\bf B378}
(1992) 175.}

\Title{\vbox{\baselineskip12pt\hbox{LA-UR-93-474}\hbox{hep-th/9302046}
}}{Constrained Dynamics of the Coupled Abelian Two-Form}
\centerline{Amitabha Lahiri\footnote{$^{\dag}$}{(lahiri@pion.lanl.gov)}}
\bigskip\centerline{Theoretical  Division  T-8}
\centerline{Los Alamos National Laboratory}
\centerline{Los Alamos, NM 87545, USA}
\vskip0.3in
\centerline{\bf Abstract}
I present the reduction of phase space of the theory of an antisymmetric
tensor potential coupled to an abelian gauge field, using Dirac's
procedure. Duality transformations on the reduced phase space are also
discussed.

\Date{01/93}


\newsec{Introduction}

The uses of an abelian antisymmetric tensor potential is manifold. It was
probably first used in the context of particle theory to describe a massless
particle of zero helicity, the notoph \refs{\op, \des}. It has subsequently
appeared in the context of the theory of fundamental strings \refs{\kr, \rwit}
and of cosmic strings and vortices \refs{\vilen, \davshe, \davis}. It has also
been used to put a topological charge (hair) on black holes \refs{\bghhs,
\abli, \cdno}, and in a mechanism for generating masses for vector bosons
\refs{\ablii, \mw} via a derivative coupling. The use of the word `abelian' in
the description means that the
antisymmetric potential $B_{\mu\nu}$ is invariant under gauge transformations
belonging to any compact gauge group, although it does have an internal
non-compact symmetry. (Nonabelian generalizations will not be
discussed here.) It is also known that the free $B_{\mu\nu}$ field is
dynamically `dual' to a scalar, quantum mechanically \refs{\burg, \kyl} as well
as classically.

In this letter, I apply Dirac's procedure \dirac\ for reduction of phase space
to the
theory of an abelian antisymmetric tensor potential coupled to an abelian gauge
field. The analysis has a couple of important differences with the
corresponding analysis for the free $B_{\mu\nu}$, which has been known for a
long time \kaul. One important point is that in the coupled theory the gauge
field $A_\mu$ provides a current for $B_{\mu\nu}$, and vice versa. As a result,
choosing a gauge can be tricky. Another difference from the free theory (and an
intriguing feature by itself) is that a duality transformation in the reduced
phase space leads to very unusual \pb s between the transformed fields and
momenta.

The theory under consideration is given by the action
\eqn\abfi{S = \int d^4x\big(- \sixth H_\mnl H^\mnl - \fourth
F_{\mu\nu}F^{\mu\nu} + {m\over
2}\epsilon^{\mnl\rho}B_{\mu\nu}F_{\lambda\rho}\big),}
where $F_{\mu\nu} = \del_{[\mu}A_{\nu} = \del_\mu A_\nu - \del_\nu A_\mu$ and
$H_{\mnl} =
\del_\mu B_{\nu\lambda} + \del_\nu B_{\lambda\mu} + \del_\lambda
B_{\mu\nu}$.  My convention in this paper will be $\epsilon^{0ijk} =
+\epsilon^{ijk}\equiv +\epsilon_{ijk}$, the metric is flat, $g_{\mu\nu} =
diag(- + + +)$.  (Thus $\epsilon_{0ijk} = -\epsilon_{ijk}$.) Then the last term
in the Lagrangian may be written as
$m\epsilon_{ijk}(B_{0i}F_{jk} + B_{ij}F_{0k})$. This action remains
invariant under the independent gauge transformations
\eqn\gauge{\eqalign{&A_\mu \to A_\mu + \del_\mu\chi, \qquad B_{\mu\nu} \to
B_{\mu\nu},\hfil\cr \hskip-1.5in{\rm and}\hskip1.5in &A_\mu \to A_\mu,\qquad
B_{\mu\nu} \to B_{\mu\nu} + \del_{[\mu}\Lambda_{\nu]}.}}
The Euler-Lagrange equations for $A_\mu$ and $B_{\mu\nu}$ are readily derived,
\eqn\aele{{\delta S\over \delta A_\mu} = \del_\nu F^{\mu\nu} - {m\over 3}
\epsilon^{\mnl\rho}H_{\nu\lambda\rho} = 0,}
and
\eqn\bele{{\delta S\over \delta B_{\mu\nu}} = \del_\lambda H^\mnl + {m\over
2}  \epsilon^{\mnl\rho}F_{\lambda\rho} = 0.}
Here my convention is $\delta B_{\mu\nu}/\delta B_{\rho\lambda} =
\half(g_{\mu\rho}g_{\nu\lambda} - g_{\mu\lambda}g_{\nu\rho})$. Contracting
\aele\ with $\epsilon_{\mnl\rho}$, substituting the resulting expression
for $H_\mnl$ into \bele\ and using the Jacobi identity $\del_{[\mu}
F_{\nu\lambda]} = 0$, one finally arrives at the wave equation
\eqn\awave{( \wave - 2m^2)F_{\mu\nu} = 0.}
This is the equation for a free massive object of mass
$\sqrt2m$. As an aside, the action discussed in \ablii\ has a factor of
$-{1\over 12}$ instead of the $-\sixth$ and ${m\over 4}$ instead of
${m\over 2}$. That produces a mass $m$ for the gauge field. The choice of
factors in
\abfi\ is an arbitrary aesthetic choice, related to the choice of the
factor of $\half$ in the definition of $\delta B_{\mu\nu}/\delta
B_{\rho\lambda}$. One may regain the so-called \kyl\ `standard' normalization
by replacing
$B_{\mu\nu}$ by $B_{\mu\nu}/\sqrt2$, $m$ by $m/\sqrt2$, and by removing the
$\half$ in the definition of $\delta B_{\mu\nu}/\delta B_{\rho\lambda}$.

\newsec{Constraints of the Theory}

The momenta conjugate to $A_i$ and $B_{ij}$ are respectively
\eqn\abfii{\Pi_i = {\delta\CL\over\delta\dot A_i} = F_{0i} +
m\epsilon_{ijk}B_{jk};\qquad \Pi_{ij} = {\delta\CL\over\delta\dot B_{ij}} =
H_{0ij}.}
The canonical Hamiltonian is then
\eqn\abfiii{\eqalign{\CH_C &= \dot A_i\Pi_i + \dot B_{ij}\Pi_{ij} - \CL\hfil\cr
&= \Pi_i\Pi_i - m\epsilon_{ijk}B_{jk}\Pi_i + \del_iA_0\Pi_i +
\Pi_{ij}\Pi_{ij} - (\del_jB_{0i} - \del_iB_{0j})\Pi_{ij} - \CL\hfil\cr &=
\half\Pi_i\Pi_i + \half\Pi_{ij}\Pi_{ij} + \sixth H_{ijk}H_{ijk} +
m^2B_{ij}B_{ij} + \fourth F_{ij}F_{ij} - m\epsilon_{ijk}\Pi_iB_{jk}\hfil\cr
& \qquad\qquad\qquad + \del_iA_0\Pi_i - (\del_jB_{0i} -
\del_iB_{0j})\Pi_{ij} - m\epsilon_{ijk}B_{0i}F_{jk}.\hfil\cr}}
The primary constraints of this system are
\eqn\apri{\Pi_0 \weakly 0,\qquad \Pi_{0i} \weakly 0.}
 So I have to add Lagrange multiples of these constraints to the canonical
Hamiltonian and integrate over all space to get a Hamiltonian
\eqn\abfv{H_0 = \int d^3x\big(\CH_C + v(\bx)\Pi_0(\bx) +
v_i(\bx)\Pi_{0i}(\bx)\big).}

The canonical \pb s are
\eqn\canpb{\eqalign{\PB{A_\mu(\bx)}{\Pi_\nu(\by)} &=
g_{\mu\nu}\delta^3({\bf x} - {\bf y}),\hfil\cr
\PB{B_{\mu\nu}({\bf x})}{\Pi_{\rho\lambda}({\bf y})} &=
\half(g_{\mu\rho}g_{\nu\lambda} - g_{\mu\lambda}g_{\nu\rho})\delta^3({\bf x}
- {\bf y}),\hfil\cr}}
all the other brackets being zero. Here and everywhere else, it is understood
that brackets are taken at
equal times. In addition, I will not specify the spatial dependence where
it is obvious. Then the compatibility conditions for the primary
constraints \apri\ are given by their \pb s with the Hamiltonian $H_0$,
\eqn\asec{\eqalign{\dot\Pi_0 = \PB{\Pi_0}{H_0} &= -\del_i\Pi_i \weakly 0;
\hfil\cr \dot\Pi_{0i} = \PB{\Pi_{0i}}{H_0} &= \del_j\Pi_{ij} -
{m\over 2}\eijk F_{jk} \weakly 0.\hfil\cr}}
As is obvious, these are respectively the 0-th and $[0i]$-th components of
the Euler-Lagrange equations. These are the secondary constraints of the
system, and one can read off that the \pb s of \asec\ with \apri\
vanish. Therefore these are all first-class constraints.

Now Lagrange multiples of these secondary constraints are added to $H_0$ to
get the unconstrained Hamiltonian
\eqn\abfviii{H = H_0 + \int d^3x\big(w(\bx)\del_i\Pi_i +
w_i(\bx)(\del_j\Pi_{ij} - {m\over 2}\eijk F_{jk})\big).}
The Lagrange multipliers $v, v_i, w$ and $w_i$ may be computed from the \pb
s of the fields with this Hamiltonian,
\eqn\abfix{\eqalign{\dot A_0 &= \PB{A_0}{H} = -v\hfil\cr
 \dot A_i &= \PB{A_i}{H} = \Pi_i + \del_iA_0 - \del_iw - m\eijk
B_{jk}\hfil\cr \dot B_{0i} &= \PB{B_{0i}}{H} = -\half v_i \hfil\cr
\dot B_{ij} &= \PB{B_{ij}}{H} = \Pi_{ij} - \del_jB_{0i} + \del_iB_{0j} -
\half(\del_jw_i - \del_iw_j).\hfil\cr}}
Therefore one may set $v= -\dot A_0, v_i = -2\dot B_{0i}, w = 0$, and $w_i
= 0$. (Actually, instead of $w = 0$ and $w_i = 0$ it is sufficient to set
$\del_iw = 0$ and $\del_{[i}w_{j]} = 0$. The resulting Hamiltonian then
differs from the one below by boundary terms corresponding to the `zero
modes' of the system. I will neglect these boundary terms in what follows.
(For an account of how the boundary terms affect the vacuum structure of
the quantum theory, and related topological issues, see \refs{\tja, \bal}.)
Finally, one may compute the \pb\ of the secondary constraints with the
Hamiltonian, $\PB{\del_i\Pi_i}{H} \weakly \del_i\del_jF_{ij} = 0$, while
\eqn\abfx{\eqalign{\PB{\del_i\Pi_{ij} - {m\over 2}\eijk F_{jk}}{H} &\weakly
2m^2\del_jB_{kl}\half(\delta_{ik}\delta_{jl} - \delta_{il}\delta_{jk}) -
m\eklm\del_j\Pi_k\half(\delta_{il}\delta_{jm} -
\delta_{im}\delta_{jl})\hfil\cr &\qquad\qquad \ + m\eijk\del_j\Pi_k -
m^2\eijk\eklm\del_jB_{lm} = 0.}}
Therefore there are no tertiary constraints, and \apri\ and \asec\ form a
complete set of first-class constraints.

\newsec{Constrained Hamiltonian}

In order to compute Dirac brackets for the system, I shall fix the gauge
by choosing `gauge-fixing' constraints with non-vanishing \pb s with the
primary and secondary constraints \apri\ and \asec. The $0$-th component of
the Euler-Lagrange equations for \aele\ reads
\eqn\aelea{- \del_0\del_iA_i + \nabla^2A_0 - {m\over3}\eijk H_{ijk} = 0.}
It follows from this equation that the gauge fixing constraints $A_0
\weakly 0$ and $\del_iA_i \weakly 0$ cannot both be imposed (unless $m =
0$). I will choose `radiation gauge' constraints
\eqn\aradg{\del_iA_i \weakly 0 \qquad{\rm and}\qquad A_0(\bx) +
{m\over3}\int{d^3y\over4\pi|\bx - \by|}\eijk H_{ijk}(\by) \weakly 0.}
This gauge-fixing can be done by performing the following gauge
transformations, first
\eqn\agaugea{A_\mu \to A'_\mu(\bx) = A_\mu(\bx) -
\del_\mu\int_0^{x_0}dt\bigg[A_0(\bx, t) +
{m\over3}\int{d^3y\over4\pi|\bx - \by|}\eijk H_{ijk}(\by, t)\bigg], }
so that
\eqn\aone{A'_0 + {m\over3}\int{d^3y\over4\pi|\bx - \by|}\eijk H_{ijk}(\by)
= 0,}
and \aelea\ becomes $\del_0\del_iA'_i = 0$, where I have used
$\nabla_x^2\big({1\over4\pi|\bx - \by|}\big) = -\delta^3(\bx - \by).$ Then
another gauge transformation is made,
\eqn\agaugeb{A'_\mu \to A''_\mu(\bx) = A'_\mu(\bx) +
\del_\mu\int{d^3y\over4\pi|\bx - \by|}\del^y_iA'_i(\by).}
Then $\del_iA''_i = 0$ and $A''_0 = A'_0$, where I have used $\del_0\del_iA'_i
= 0$.

Similarly, the $0i$-th component of the Euler-Lagrange equations \bele\
reads
\eqn\belea{-\del_0\del_jB_{ij} - \nabla^2B_{0i} + \del_i\del_jB_{0j} +
{m\over2}\eijk F_{jk} = 0.}
Again, $B_{0i} \weakly 0$ and $\del_jB_{ij} \weakly 0$ cannot both be
imposed unless $m = 0$. I shall choose what may be called `radiation gauge'
constraints for the $B_{\mu\nu}$ field,
\eqn\bradg{\del_jB_{ij} \weakly 0 \qquad{\rm and}\qquad B_{0i} +
{m\over2}\int{d^3y\over4\pi|\bx - \by|}\eijk F_{jk}(\by) \weakly 0.}
These choices can be made via the following gauge transformations (see
\gauge), first
\eqn\bgaugea{B_{\mu\nu} \to B'_{\mu\nu}(\bx) = B_{\mu\nu}(\bx) +
\del_{[\mu}\int_0^{x_0}dt\bigg[B_{\nu]0}(\bx, t) -
{m\over2}\int{d^3y\over4\pi|\bx - \by|}\epsilon_{\nu]0jk}F_{jk}(\by, t)\bigg],}
so that
\eqn\bone{B'_{0i} + {m\over2}\int{d^3y\over4\pi|\bx - \by|}\eijk F_{jk}(\by)
= 0,}
and \belea\ is now $\del_0\del_jB'_{ij} = 0$. Then
the radiation gauge is achieved via another gauge transformation
\eqn\bgaugeb{B'_{\mu\nu} \to B''_{\mu\nu} = B'_{\mu\nu} -
\del_\mu\int{d^3y\over4\pi|\bx -
\by|}\del^y_kB'_{\nu k} + \del_\nu\int{d^3y\over4\pi|\bx -
\by|}\del_k^yB'_{\mu k}.}
In this gauge, $\del_jB''_{ij} = 0$ and $B''_{0i} = B'_{0i}$.

Now I can drop all primes and write down the full set of second-class
constraints,
\eqn\aset{\eqalign{\phi_1&\equiv A_0(\bx) + {m\over3}\int{d^3y\over4\pi|\bx -
\by|}\eijk H_{ijk}(\by) \weakly 0,\hfil\cr
\phi_2&\equiv\Pi_0\weakly 0,\hfil\cr
\phi_3&\equiv\del_iA_i\weakly 0, \hfil\cr
\phi_4&\equiv\del_i\Pi_i\weakly 0,\hfil\cr
\phi_{5i}&\equiv B_{0i} + {m\over2}\int{d^3y\over4\pi|\bx - \by|}\eijk
F_{jk}(\by)\weakly 0,\hfil\cr
\phi_{6i}&\equiv\Pi_{0i}\weakly 0,\hfil\cr
\phi_{7i}&\equiv\del_jB_{ij}\weakly 0,\hfil\cr
\phi_{8i}&\equiv\del_j\Pi_{ij} - {m\over 2}\eijk F_{jk}\weakly 0.\hfil\cr}}
The matrix of \pb s of these constraints are easily computed from the canonical
\pb s \canpb. In particular, the matrix of \pb s of the constraints $\phi_{7i}$
and $\phi_{8i}$ is
\eqn\matij{\eqalign{\PB{\phi_{7i}(\bx)}{\phi_{8j}(\by)} &= \del_k^x\del_l^y
\half(g_{ij}g_{kl} - g_{il}g_{jk})\delta^3(\bx - \by)\hfil\cr
&= -\half(\nabla^2_xg_{ij} - \del_i^x\del_j^x)\delta^3(\bx - \by).\hfil\cr}}
As is well-known, the right-hand side of this equation is non-invertible.
It follows that the constraints $\phi_{7i}$ and $\phi_{8i}$ are degenerate.
This degeneracy is due to the fact that there is a propagating mode of the
$B_{\mu\nu}$ field. In four dimensions, the antisymmetric $B_{\mu\nu}$ has
six independent components. On the other hand it appears as if there are
six first-class constraints \apri\ and \asec\ on these components. These,
however, are not independent constraints and their interdependence is
carried through after gauge fixing into the second class constraints \aset.
The non-invertibility of the matrix of \pb s shows that \aset\ is
overcounting the number of second-class constraints.  In order to compute
the Dirac brackets of the fields, one needs the inverse of the matrix of
the \pb s of the second-class constraints. Because of the non-invertibility
of \matij\ one has to resort to a trick.  This trick is to introduce a
parameter $\eta$, and redefine the relevant \pb s,
\eqn\trick{\PB{B_{\mu\nu}(\bx)}{\Pi_{\rho\lambda}(\by)} =
\half(g_{\mu\rho}g_{\nu\lambda} - (1 - {1\over\eta})g_{\mu\lambda}g_{\nu\rho})
\delta^3(\bx - \by).}
Then the \pb s of $\phi_{7i}$ and $\phi_{8i}$ are given by
\eqn\trickpb{\PB{\phi_{7i}(\bx)}{\phi_{8j}(\by)} = -\half(\nabla^2_xg_{ij} -
(1 - {1\over\eta})\del_i^x\del_j^x)\delta^3(\bx - \by) =: - \half
G_{ij}(\bx)\delta^3(\bx - \by).}
I will compute the Dirac brackets as usual, then take the limit
$\eta\to\infty$ at the end. The matrix of \pb s of the constraints \aset\
is now given by
\eqn\cmat{{\CC}_{\alpha\beta}({\bf z}, {\bf w}) =
\pmatrix{&0&-1&0&0&0&0&0&0\cr&1&0&0&0&0&0&0&0\cr&0&0&0&-\nabla^2_z&0&0&0&0\cr
&0&0&\nabla^2_z&0&0&0&0&0\cr&0&0&0&0&0&-\half g_{ij}&0&0\cr &0&0&0&0&\half
g_{ij}&0&0&0\cr&0&0&0&0&0&0&0&\kern-10pt-\half G_{ij}({\bf
z})\cr&0&0&0&0&0&0&\kern-10pt\half G_{ij}({\bf z})&0\cr}\delta^3({\bf z} -
{\bf w}),}
where $\alpha, \beta = 1, \cdots, 8.$The inverse of this matrix is easily
computed to be
\eqn\invmat{{\CC}^{-1}_{\alpha\beta}({\bf z}, {\bf w}) =
\pmatrix{&0&1&0&0&0&0&0&0\cr&-1&0&0&0&0&0&0&0\cr&0&0&0&{1\over\nabla^2_z}
&0&0&0&0\cr&0&0&-{1\over\nabla^2_z}&0&0&0&0&0\cr&0&0&0&0&0&
2 g_{ij}&0&0\cr&0&0&0&0&-2 g_{ij}&0&0&0\cr&0&0&0&0&0&0&0&\kern-10pt
2G^{-1}_{ij}({\bf z})\cr&0&0&0&0&0&0&\kern-10pt-2G^{-1}_{ij}({\bf
z})&0\cr}\delta^3({\bf z} - {\bf w}),}
where $\displaystyle{1\over\nabla^2_z}\delta^3({\bf z} - {\bf w}) = -
{1\over 4\pi{|{\bf z} - {\bf w}|}}$, and $G^{-1}_{ij}$ is a formal inverse,
\eqn\ginv{G^{-1}_{ij}({\bf z}) =  \bigg(g_{ij} - (1 -
\eta){\del^z_i\del^z_j\over\nabla^2_z}\bigg){1\over\nabla^2_z}.}
According to Dirac's prescription, the brackets among two objects $\CA$
and $\CB$ may now be calculated, consistent with setting the constraints
\aset\ strongly to zero, by the formula
\eqn\dbra{\DB{\CA({\bf x})}{\CB({\bf y})} = \PB{\CA({\bf x})}{\CB({\bf
y})} - \int d^3zd^3w \PB{\CA({\bf x})}{\phi_\alpha({\bf
z})}{\CC}^{-1}_{\alpha\beta}({\bf z} - {\bf w})\PB{\phi_\beta({\bf
w})}{\CB({\bf y})},}
where the brackets on the right hand side of the equation are the canonical
\pb s given in \canpb. The Dirac brackets of $A_\mu$ and $\Pi_\nu$ can be
easily computed,
\eqn\abra{\DB{A_\mu({\bf x})}{\Pi_\nu({\bf y})} = (g_{\mu\nu} +
g_{0\mu}g_{0\nu})\delta^3(\bx - \by) - \del^x_\mu\del^y_\nu{1\over4\pi|{\bf
x} - {\bf y}|}.}

For the calculation of the Dirac brackets of $B_{\mu\nu}$ and
$\Pi_{\rho\lambda}$, one calculates with ${\CC}^{-1}_{\alpha\beta}$ as
given in \invmat\ and then takes $\eta\to\infty$. Thus
\eqn\bbra{\eqalign{\DB{B_{\mu\nu}({\bf x})}{\Pi_{\rho\lambda}({\bf y}
)}^\eta = &\ \half(g_{\mu\rho}g_{\nu\lambda} -
g_{\mu\lambda}g_{\nu\rho})\delta^3({\bf x}
- {\bf y}) \hfil\cr & +\half
\Big[g_{\mu0}g_{0\rho}g_{i\nu}g_{i\lambda} - (1 -
{1\over\eta})g_{\nu0}g_{0\rho}g_{i\mu}g_{i\lambda} \hfil\cr &-(1 -
{1\over\eta})g_{\mu0}g_{0\lambda}g_{i\nu}g_{i\rho} + (1 -
{1\over\eta})^2g_{\nu0}g_{0\lambda}g_{i\mu}g_{i\rho}\Big]\delta^3({\bf x}
- {\bf y})\hfil\cr & -\half\Big[(g_{\mu\rho} +
g_{\mu0}g_{0\rho})\del_\nu^x\del_\lambda^y - (1 -
{1\over\eta})(g_{\mu\lambda} + g_{\mu0}g_{0\lambda})\del_\nu^x\del_\rho^y
\hfil\cr &- (1 - {1\over\eta})(g_{\nu\rho} +
g_{\nu0}g_{0\rho})\del_\mu^x\del_\lambda^y \hfil\cr &+ (1 -
{1\over\eta})^2(g_{\nu\lambda} +
g_{\nu0}g_{0\lambda})\del_\mu^x\del_\rho^y\Big]{1\over4\pi|{\bf x} - {\bf
y}|}\hfil\cr & +
{1\over\eta^2}\del_\mu^x\del_\nu^x\del_\rho^y\del_\lambda^y\big({{1 -
\eta}\over\nabla^2_x}\big){1\over4\pi|{\bf x} - {\bf y}|}\hfill ,}}
for which one takes $\eta\to\infty$ and obtains the Dirac brackets for
$B_{\mu\nu}$ and $\Pi_{\rho\lambda}$,
\eqn\bdbra{\eqalign{\DB{B_{\mu\nu}({\bf x})}{\Pi_{\rho\lambda}({\bf y})}
=&\
\half(g_{\mu\rho}g_{\nu\lambda} - g_{\mu\lambda}g_{\nu\rho})\delta^3({\bf
x} - {\bf y})\hfil\cr & +\half(g_{\mu0}g_{0\rho}g_{\nu\lambda} -
g_{\nu0}g_{0\rho}g_{\mu\lambda} - g_{\mu0}g_{0\lambda}g_{\nu\rho} +
g_{\nu0}g_{0\lambda}g_{\mu\rho})\delta^3({\bf x} - {\bf y})\hfil\cr
& - \half\Big[(g_{\mu\rho} + g_{\mu0}g_{0\rho})\del_\nu^x\del_\lambda^y
- (g_{\mu\lambda} + g_{\mu0}g_{0\lambda})\del_\nu^x\del_\rho^y \hfil\cr
& - (g_{\nu\rho} + g_{\nu0}g_{0\rho})\del_\mu^x\del_\lambda^y
+ (g_{\nu\lambda} + g_{\nu0}g_{0\lambda})\del_\mu^x\del_\rho^y\Big]
{1\over4\pi|{\bf x} - {\bf y}|}\ .}}
The other Dirac brackets may be computed in a similar fashion, and they are
\eqn\obra{\eqalign{\DB{A_\mu(\bx)}{\Pi_{\nu\lambda}(\by)} & =
2mg_{0\mu}\epsilon_{\nu\lambda\rho}\del_\rho^y{1\over4\pi|\bx - \by|}\hfil\cr
\DB{B_{\mu\nu}(\bx)}{\Pi_\lambda(\by)} & =
m(g_{0\nu}\epsilon_{\lambda\mu\sigma} -
g_{0\mu}\epsilon_{\lambda\nu\sigma})\del_\sigma^y{1\over4\pi|\bx -
\by|}\hfil\cr
\DB{\Pi_\mu(\bx)}{\Pi_{\nu\lambda}(\by)} & =
m(\epsilon_{\mu\lambda\sigma}\del_\sigma^x\del_\nu^y -
\epsilon_{\mu\nu\sigma}\del_\sigma^x\del_\lambda^y)
{1\over4\pi|\bx - \by|}\hfil\cr
\DB{A_\mu}{A_\nu} = \DB{A_\mu}{B_{\mu\nu}} & = \DB{B_{\mu\nu}}{B_{\rho\lambda}}
= \DB{\Pi_\mu}{\Pi_\nu} = \DB{\Pi_\mu}{\Pi_{\mu\nu}} = 0.}}
Here $\epsilon_{\mu\nu\lambda} = \eijk g_{i\mu}g_{j\nu}g_{k\lambda}$.

One may then set the constraints \aset\ strongly to zero. The fully
constrained Hamiltonian is then
\eqn\conham{H = \int d^3x\big(\half\Pi_i\Pi_i + \half\Pi_{ij}\Pi_{ij} +
\sixth H_{ijk}H_{ijk} + \fourth F_{ij}F_{ij} + m^2B_{ij}B_{ij} -
m\eijk\Pi_iB_{jk}\big),}
with the fields and the momenta obeying the Dirac brackets as given by
\abra\ -- \obra.  The linear nature of the coupling between $\Pi_i$ and
$B_{jk}$ may raise questions as to the boundedness (from below) of this
Hamiltonian. However,
\conham\ is bounded from below, since
\eqn\bound{\eqalign{H \equiv& \int d^3x\big(\half(\Pi_i - m\eijk
B_{jk})(\Pi_i - m\epsilon_{ilm}B_{lm}) + \half\Pi_{ij}\Pi_{ij} + \sixth
H_{ijk}H_{ijk} +\fourth F_{ij}F_{ij}\big)\hfil\cr =& \int d^3x(\half
F_{0i}F_{0i} + \half\Pi_{ij}\Pi_{ij} + \sixth H_{ijk}H_{ijk} +\fourth
F_{ij}F_{ij}),}}
which is a sum of squares.

The duality between the antisymmetric potential and a scalar is quite
interesting in this formalism. First, let $\Pi'_i$ be defined as $\Pi'_i :=
F_{0i} = \Pi_i - m \eijk B_{jk}$ (a non-canonical transformation). Then
from \bdbra\
and \obra, one calculates that
\eqn\fbra{\DB{\Pi'_i(\bx)}{\Pi_{jk}(\by)} = - m\eijk\delta^3 (\bx - \by).}
On the other hand, $\Pi'_i$ has the same Dirac brackets as $\Pi_i$ with all
the other variables that appear in the Hamiltonian \conham. I now drop the
primes and rewrite the Hamiltonian as
\eqn\newham{H = \int d^3x\big(\half\Pi_i\Pi_i + \half\Pi_{ij}\Pi_{ij} +
\sixth H_{ijk}H_{ijk} + \fourth F_{ij}F_{ij}\big).}
The fields are now uncoupled in the Hamiltonian but coupled through their
Dirac brackets \fbra. Even though this was not a canonical transformation,
no information has been lost. Now I make a canonical transformation from the
variables $\big(B_{\mu\nu}, \Pi_{\rho\lambda}\big)$ to the scalar variables
$\big(\phi, \Pi_\phi\big)$,
\eqn\cantrans{\phi(\bx) = {1\over \sqrt2}\eijk\int_x^\infty d\lambda
{dz_i\over d\lambda}\Pi_{jk}({\bf z}), \qquad \Pi_\phi
= {1\over \sqrt2}\eijk\del_iB_{jk}.}
The inverse transformation is also non-local,
\eqn\invtrans{\Pi_{ij} = {1\over\sqrt2}\eijk\del_k\phi, \qquad
B_{ij}(\bx) = {1\over3\sqrt2}\eijk\int_x^\infty d\lambda{dz_k\over
d\lambda}\Pi_\phi.}
The hamiltonian may now be rewritten in terms of the variables $\big(A_i,
\Pi_i, \phi, \Pi_\phi\big)$,
\eqn\phiham{H = \int d^3x\big(\half\Pi_i\Pi_i + \fourth F_{ij}F_{ij} +
 \half\Pi_\phi\Pi_\phi + \half\del_i\phi\del_i\phi\big),}
with Dirac brackets given by \abra\ (where $\Pi_i$ are the new $\Pi_i$), and
\eqn\phibra{\eqalign{\DB{\phi(\bx)}{\Pi_\phi(\by)} & = \delta^3(\bx - \by)
\hfil\cr
\DB{A_i}{\phi} = \DB{A_i}{\Pi_\phi} & = \DB{\Pi_i}{\Pi_\phi} = 0\hfil\cr
\DB{\Pi_i(\bx)}{\phi(\by)} & = m\sqrt2\int_y^\infty dz_i\delta^3(\bx
- {\bf z}).}}
While $\big(B_{\mu\nu}, \Pi_{\rho\lambda}\big)\to\big(\phi, \Pi_\phi\big)$
is a canonical transformation, it is also non-local, and therefore
incomplete. The systems
\newham\ and \phiham\ carry different topological information, as is
evident from the brackets \phibra. While the brackets in the first system
were local, the brackets now depend on a line of integration. Various
authors \refs{\davshe, \davis, \kyl, \tja, \cpw} have pointed out relations
between cosmic
strings in the system \abfi\ and local strings of the abelian Higgs model
in the symmetry broken phase. The transformations \cantrans, \invtrans\ and
the \pb s \phibra\ seem to point to yet another description of the duality
between the two systems.

\listrefs
\bye